\documentclass[aps,prl,twocolumn,showpacs,superscriptaddress]{revtex4-1}
\usepackage{amsmath}
\usepackage{epsfig}
\usepackage{color}
\usepackage{endnotes}
\usepackage{natbib}
\usepackage[colorlinks,breaklinks]{hyperref}
\usepackage{bm}% bold math
\usepackage{dcolumn}
\usepackage{graphics}% Include figure files
\usepackage{graphicx} %for inset figure?
\usepackage{tikz} %for inset figure?
\usepackage{amsmath}
\usepackage{amssymb}
\usepackage{color}
\usepackage{changes}
\usepackage{multirow}
\newcommand{\bs}[1]{\ensuremath{\boldsymbol{#1}}}

\newcommand{\be}{\begin{equation}}
\newcommand{\ee}{\end{equation}}
\newcommand{\bea}{\begin{align}}
\newcommand{\eea}{\end{align}}
\newcommand{\beqa}{\begin{eqnarray}}
\newcommand{\eeqa}{\end{eqnarray}}

\newcommand{\rvec}{\bs{r}}
\newcommand{\Rvec}{\bs{R}}

\begin{document}

\title{
{\small \hskip12cm NT@UW-18-07\\}
Short-range correlations and the charge density}

\author{Ronen Weiss}
\affiliation{The Racah Institute of Physics, The Hebrew University, 
             Jerusalem, Israel}
\author{Axel Schmidt}
\affiliation{Massachusetts Institute of Technology, Laboratory for Nuclear Science, Cambridge, MA 02139, USA}
\author{Gerald A. Miller}
\affiliation{University of Washington, Department of Physics, Seattle, WA 98195, USA}
\author{Nir Barnea}
\email{nir@phys.huji.ac.il}
\affiliation{The Racah Institute of Physics, The Hebrew University, 
             Jerusalem, Israel}

\date{\today}
\begin{abstract} 
Sophisticated high-energy and large momentum-transfer scattering experiments
combined with ab-initio calculations can
reveal the short-distance behavior of nucleon pairs in nuclei.
%such as the probability of finding short-range correlated pairs
%in different nuclei. 
On an opposite energy and resolution scale, elastic electron scattering experiments
are used to extract the charge density and charge radius of different nuclei.
%The charge density can be well described using mean field theories, i.e. without
%considering any effects of short-range correlations.
We show  that even though the charge density has no obvious  connection with 
  nuclear short-range correlations,
it can be used to extract properties of such correlations.
%like the probability of finding a pair of nucleons
%close to each other inside the nucleus.
This is accomplished by using 
  the nuclear contact formalism   to derive
a relation between the charge density and the  proton-proton nuclear contacts that 
describe the probability of two protons being at close proximity. 
%This is done by building a simple model for the
%proton-proton pair density that is verified using available ab-initio calculations.
With this relation, the values of the proton-proton contacts
are extracted for various nuclei  using only  the  nuclear 
charge density and a solution of the two-nucleon Schroedinger equation
as   inputs.
For symmetric nuclei, the proton-neutron
contacts can also be extracted from the charge density.
Good agreement is obtained with previous extractions
of the nuclear contacts. 
These results imply %leads to the surprising conclusion 
that one can predict (with reasonably good accuracy) the results of
high-energy and large momentum-transfer electron-scattering experiments
and ab-initio calculations of high momentum tails using only experimental data of elastic
scattering experiments. 
\end{abstract}

% 67.85.-d - ultracold gases
% 05.30.Fk - Fermion systems and electron gas
% 25.20.-x - Photonuclear reactions
%\pacs{67.85.-d, 05.30.Fk, 25.20.-x}

\maketitle
%=============================================================================
% Introduction
%=============================================================================
%{\it Introduction --}

Many efforts have been devoted in the last couple of decades
to the study of nuclear short-range correlations (SRCs) and 
the short-range properties of the nuclear force.
Sophisticated high-energy and large momentum-transfer
electron and proton scattering experiments
\cite{ FraSar93,Egiyan03,Egiyan06,Fomin12,Tang2003,
Piasetzky06,Subedi08,Korover14,HenSci14,Baghdasaryan10,Shneor2007},
together with ab-initio calculations
 \cite{Schiavilla07,AlvCioMor08,FeldNeff11,AlvCio13,nuclear_matter,
CioStr91,CioSim96,CioMezMor17a,CioMor17b}, were performed.
This led to a good understanding of the main properties of
nuclear SRCs. For example, the dominance of neutron-proton
pairs due to the significant nuclear tensor force was identified.
Calculations of momentum distributions of different nuclei revealed
high momentum tails similar in shape to the deuteron high momentum tail, 
showing the universal aspects of SRCs.
Nevertheless, ab-initio numerical calculations are 
limited to light and medium-size nuclei, and only recently SRCs calculations 
for $^{40}$Ca
became accessible. 
In addition,  experimental data are only available for selected nuclei.
See also recent reviews \cite{Hen_review,Cio15_review,Fomin:2017ydn}.

Significant progress in the study of SRCs was made in the field
of atomic physics when the contact theory was presented \cite{Tan08}.
A single parameter, called the contact, describing the probability
to find two atoms close to each other, was shown to be related to many other
properties of the atomic system \cite{Bra12}. Some of these relations
are intuitive, such as a relation between the contact and the high momentum tail of the momentum distribution.
Others, however, are less intuitive, such as relationships between the 
contact parameter and thermodynamic properties of the system, for instance, 
the energy of the system, its pressure, and its entropy. % and more.
See also Ref.~\cite{Miller:2017thf} for improvements in a few of these relations.

The contact formalism was recently generalized to nuclear systems 
\cite{WeiBazBar15,WeiBazBar15a,WeiBar17,WeiHen18}.
The nuclear contacts were defined, and were shown to be related
to many different nuclear quantities, such as the two-nucleon density \cite{WeiHen18},
high momentum tails \cite{WeiBazBar15a,WeiHen18,AlvCio16},
the Coulomb sum-rule \cite{WeiPazBar}, the Levinger constant \cite{WeiBazBar15,WeiBazBar16},
and electron scattering experiments \cite{WeiBar18}. However, the
corresponding connection between the nuclear contacts and the low-energy or 
thermodynamic nuclear properties was not discovered. 

Here, we use the nuclear contact formalism
to show that nuclear SRCs and the nuclear charge-density,
closely related to
% i.e.
the one-body proton density, do indeed have  direct connection.  
This is surprising because
the charge density of a given nucleus is measured in elastic
scattering experiments that are much simpler than the high-energy experiments
devoted to the study of SRCs. The charge density and charge radius of
nuclei can be explained using mean field theories, i.e.
without an explicit need for nuclear SRCs. Experimental results of the charge
density are available for many nuclei, see e.g. Ref. \cite{charge_denisty_parameters}.
Therefore, it  might  seem that the charge density and SRCs are two
unrelated aspects of nuclear systems.
To connect these two entities we use both 
the charge density and the contact formalism to build  
a simple model for the proton-proton pair density $\rho_{pp}(\rvec)$,
i.e. for the probability of finding two protons separated by 
a distance $\rvec$. 
Assuming only the continuity of the proton-proton pair density, we obtain 
 a direct relation between the nuclear contacts and the
charge density. Since the charge density of many nuclei is known experimentally,
this new relation can be used to understand the properties of nuclear
SRCs of nuclei that are not reachable by ab-initio calculations or not yet
studied experimentally in SRC experiments.

%=============================================================================
% the nuclear contact formalism
%=============================================================================
%{\it the nuclear contact formalism --}

The two main building blocks of the nuclear contact formalism are the contacts and 
the universal functions $\varphi_{ij}^\alpha(r)$, with the index ${ij}$ representing  the three possible pairs
of nucleons: proton-proton ($pp$), proton-neutron ($pn$) and neutron-neutron ($nn$) and 
$\alpha$ the quantum numbers of the pair.  The universal functions describe the motion of a pair
of nucleons being close to each other inside the nucleus, interacting mostly
with each other, and not with the rest of the nucleons in the system.
They can be simply calculated by solving the two-nucleon Schroedinger
equation for zero-energy with a given nucleon-nucleon potential.
The two  most significant  channels are \cite{WeiHen18}: 
the spin-zero s-wave channel, occupied by all three kinds of pairs
($pp$, $pn$ and $nn$) and denoted by $\alpha=0$,
and the spin-one deuteron channel (s-wave and d-wave coupled), occupied only
by $pn$ pairs and denoted by $\alpha=1$.
Here, the nucleon-nucleon AV18 potential \cite{av18} is used for the calculation of
the universal functions.
$\varphi_{ij}^\alpha(r)$ are normalized such that
$\int_{k_F}^\infty |\tilde{\varphi}_{ij}^\alpha(\bs{k})|^2 d\bs{k}/(2\pi)^3=1$,
where $\tilde{\varphi}_{ij}^\alpha(\bs{k})$ is the Fourier transform of $\varphi_{ij}^\alpha(r)$,
and $k_F=1.3$ fm$^{-1}$.
%The conventions of Ref. \cite{WeiHen18} are used to specify the normalization of $\varphi_{ij}^\alpha(r)$.

The nuclear contacts are generally matrices denoted by $C_{ij}^{\alpha\beta}$,
but we will focus here only on the diagonal elements $C_{ij}^\alpha$,
which are proportional to the probability of finding an $ij$ pair in the channel $\alpha$
close to each other in the nucleus.
The values of the contacts are nucleus-dependent, while the two-body functions
$\varphi_{ij}^\alpha(r)$ are identical for all nuclei.
As mentioned above, several relations connecting these nuclear contacts and
different nuclear quantities and reactions have been derived. Thus,
given  the values of the contacts, different
experimental and numerical results can be described.
Recently,  the values of the contacts for
several $A \leq 40$ nuclei  have been extracted \cite{WeiHen18}, from available Variational Monte 
Carlo (VMC) calculations \cite{WirSchPie14,Wiringa_CVMC}. Obtaining the values of the contacts is 
 still a challenge for heavier nuclei.
We note that the description of SRCs using the contacts and the universal functions
is based on the asymptotic factorization of the total wave function \cite{WeiBazBar15a}.
The wave function of the VMC method is built as a product
of Jastrow correlation functions \cite{WirSchPie14}, resembling the assumed factorization. 
Therefore, the asymptotic factorization should be further investigated
using other ab-initio methods, but this goes beyond the scope of this work.

%=============================================================================
% the two body density
%=============================================================================
%{\it proton-proton density --}
We now connect nuclear SRCs with 
the charge density using  a simple description of the two-body $pp$  pair-density.
The $pp$ pair-density $\rho_{pp}(\rvec)$ describes the probability
to find a $pp$ pair at a relative distance $r$ in a given nucleus,
and is normalized to the total number
of $pp$ pairs, i.e. $\int d\rvec\, \rho_{pp}(\rvec) =Z(Z-1)/2$.
For small distances, this density is clearly related to short-range correlations,
and can be expressed using the nuclear contacts \cite{WeiHen18}
\be \label{contact_rel}
  \rho_{pp}(\rvec) = C_{pp}^0 |\varphi_{pp}^0(\rvec)|^2.
\ee
Previous work \cite{WeiHen18} found that
this relation holds for $r<r_0 \approx 0.9$ fm for 
nuclei with $A \leq 40$.

For large separation distances, we expect that no correlations
will be relevant and thus the two-body $pp$ pair-density
can be written using the  
one-body point-proton density
%charge density
 $\rho_p(\rvec)$
\cite{correlation_func}:
\be \label{UC_rho_pp}
  \rho_{pp}(\rvec) \propto 
  \rho_{pp}^{UC}({\rvec}) \equiv \int d\bs{R}\, \rho_p(\bs{R}+\bs{r}/2)
\rho_p(\bs{R}-\bs{r}/2),
\ee
integrating over all possible locations of the center-of-mass $\bs{R}$ of the
$pp$ pair. This asymptotic behavior   does not  account  for the
fermionic nature of the $pp$ pair. To understand its effect,
we examine  the Fermi-gas model for infinite nuclear matter having a constant
 proton density $\rho_p$.
In this model, the probability to find two protons at positions $\bs{r}_1$
and $\bs{r}_2$ is given by
\be \label{Fermi_correction}
\rho_{pp}(\bs{r}_1,\bs{r}_2) = 
%\frac{1}{2} \rho_p^2 \left[1-\frac{1}{2} C^2 \left( k_F^p|\bs{r}_1-\bs{r}_2| \right) \right]
\frac{1}{2} \rho_p^2 \left[1-\frac{1}{2} \left( \frac{3 j_1(k_F^p r)}
{k_F^p r} \right)^2  \right],
\ee
where $j_1$ is a spherical Bessel function,
$r =  |\bs{r}_1-\bs{r}_2|$ 
and $k_F^p$ is the proton Fermi momentum.
Based on this expression, and integrating over the
center of mass of the pair $\bs{R} = \bs{r}_1+\bs{r}_2$, we expect that
the $pp$ density of finite nuclei at large distances will obey
\be \label{large_distances_rho_pp}
\rho_{pp}(r) \xrightarrow[r \rightarrow \infty]{}
\rho_{pp}^F(r)
\equiv {\cal N} \rho_{pp}^{UC}(r)
 \left[1-\frac{1}{2}\left( \frac{3 j_1(k_F^p r)}
{k_F^p r} \right)^2 \right].
\ee
Here, ${\cal N}$ is a normalization factor, fixing the normalization 
of $\rho_{pp}^F(r)$ to the number of $pp$ pairs.
This provides an asymptotic expression for the $pp$
density that can be calculated directly from the
one-body point-proton density. % charge density.
The charge density, measured in elastic scattering experiments,
is slightly different than the point-proton density, due to the structure
of protons and neutrons and their internal charge distribution.
Nevertheless, for medium-size and heavy nuclei
this difference becomes small, and the experimental charge distribution
can be used in Eq. \eqref{UC_rho_pp} to a good approximation.
In addition, since Eq. \eqref{large_distances_rho_pp} is based on
the nuclear matter expression, we might not expect it to hold
for the light nuclei.
Shell model calculations for $^{16}$O using harmonic oscillator orbitals 
(with $\sqrt{\hbar/m\omega}=1.79$ fm, following Ref. \cite{Negele}) agree with the
plane-wave nuclear matter correction
(used in Eqs. \eqref{Fermi_correction} and \eqref{large_distances_rho_pp}).
For $k_F^p=0.9$ fm$^{-1}$
less than 2\% difference is seen for $r<4$ fm,
and for a more realistic value of $k_F^p=1.05$ fm$^{-1}$ an agreement 
with 10\% accuracy is obtained for the same range.

If SRCs were not significant in nuclear systems,
then $\rho_{pp}^F(r)$ might have been a good approximation
for the exact $\rho_{pp}(r)$ for all $r$. Thus,
we can expect that the asymptotic expression of Eq.
\eqref{large_distances_rho_pp} will hold for $r \gtrsim r_0$,
because SRCs are significant for $r \lesssim r_0$.

\begin{figure}\begin{center}
\includegraphics[width=8.6 cm]{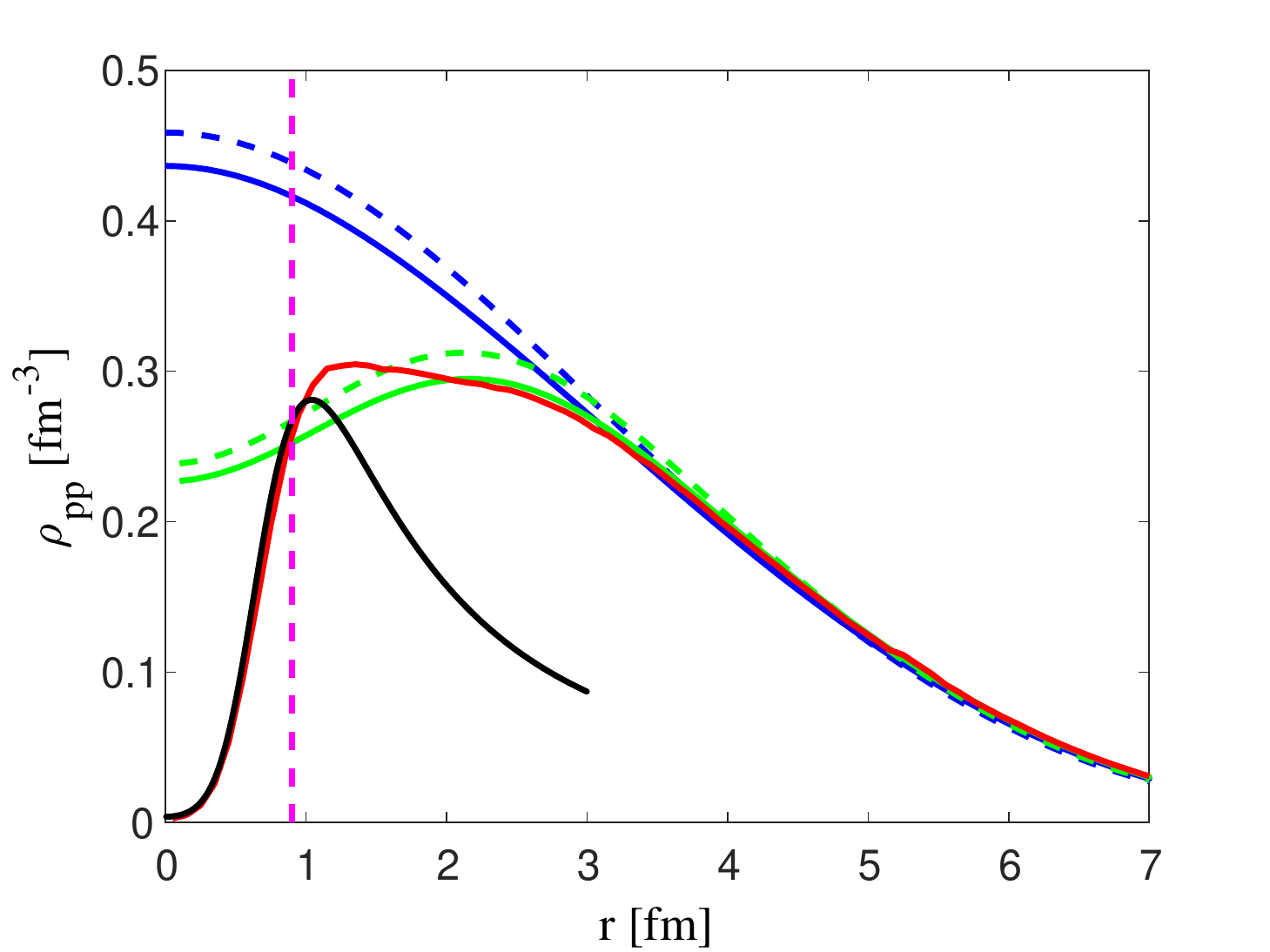}
\caption{\label{40Ca}
The $pp$ density for $^{40}$Ca. 
The red line shows the full $pp$ density $\rho_{pp}$ of the VMC
calculations \cite{Wiringa_CVMC}. The blue lines show the calculated $\rho_{pp}^{UC}$,
based on Eq. \eqref{UC_rho_pp},
using the one-body point-proton density of the VMC calculations
(solid) or the experimental charge density (dashed), using the 
three-parameter Fermi model \cite{charge_denisty_parameters}.
The green lines show the corresponding
uncorrelated $pp$ density with the Fermi-statistic correction,
using Eq. \eqref{large_distances_rho_pp}.
The black line shows the contact expression
for the $pp$ density, using the AV18 potential and the contact
value extracted in Ref. \cite{WeiHen18} by fitting
to the VMC data in coordinate space.
The blue, green and red lines are all normalized to the number of $pp$ pairs.
The vertical magenta line shows the location of $r=0.9$ fm.
}
\end{center}\end{figure}

We now have expressions for both small-distance
and large-distance asymptotics of $\rho_{pp}(r)$ that can be 
compared  to results of available VMC
numerical calculations \cite{WirSchPie14,Wiringa_CVMC},
calculated using the AV18 \cite{av18} and UX \cite{ubx} potentials.
 The results for $^{40}$Ca are presented in Fig. \ref{40Ca}.
First observe that  the uncorrelated $\rho_{pp}^{UC}$,
calculated using either the VMC point-proton density
%charge distribution
or experimental
charge distribution \cite{charge_denisty_parameters},
coincides with the VMC $pp$ density for large distances
($r \gtrsim 3$ fm). Then note that 
the uncorrelated $pp$ density including the Fermi statistic
$\rho_{pp}^F$ describes (as expected) the full $\rho_{pp}$
density for $r \gtrsim r_0$ reasonably well. For
smaller separations the contact relation,
Eq. \eqref{contact_rel}, using the AV18 potential, is seen to  agree with the VMC calculations
for $r \lesssim r_0$.
Most importantly, one can see that around $r_0 \approx 0.9$ fm
both the contact and the $\rho_{pp}^F$ expressions seem to describe the value of the full $\rho_{pp}$ reasonably
well.

The Fermi momentum is calculated  via
its relation to the proton density. For infinite nuclear matter 
$k_F^p = (3\pi^2 \rho_p)^{1/3}.$
For finite nuclei, $\rho_p$ depends on the location. 
In the local density approximation,
the Fermi momentum at the
pair's center of mass $\bs{R}$ is given by
\be 
  k_F^p(\Rvec) = (3\pi^2 \rho_p(\Rvec))^{1/3}.
\ee
We can use this expression of the  proton Fermi momentum 
$k_F^p(\Rvec)$ for evaluating Eq.~\eqref{Fermi_correction},
or instead we can use the weighted average value
\be \label{kF_averaged}
  k_F^p = \frac{\int d\rvec\, k_F^p(\rvec) \rho_p(\rvec)}{\int d\rvec\, \rho_p(\rvec)}.
\ee
In the calculations presented in Fig. \ref{40Ca}, we have
used this last relation  
resulting in a numerical value of $k_F^p \approx 1.05$ fm$^{-1}$ for $^{40}$Ca.
Another possible choice for $k_F^p$ is to use the value of 
$\rho_p(r)$ at the center of the
nucleus, i.e. $r=0$, 
\be \label{kF_central}
  k_F^p= (3\pi^2 \rho_p(0))^{1/3}.
\ee
In any case, we only need to know the proton density $\rho_p(r)$ to obtain $k_F^p$.
We will use below Eq. \eqref{kF_averaged} for the calculation of
the Fermi momentum. We note that the following results are not sensitive to the
exact value of $k_F^p$, and are almost unchanged if Eq. \eqref{kF_central}
is used instead of Eq. \eqref{kF_averaged}.

We now use  these results to extract the value of the $pp$ contact for
any nucleus using only its charge distribution.
 Since $\rho_{pp}$ should be well described by $\rho_{pp}^F$ for $r>r_0\approx 0.9$ fm
and by the contact expression for $r<r_0$, we can extract the value of $C_{pp}^0$
by only requiring the continuity of $\rho_{pp}$ at $r=r_0$. 
This gives   the relation
\be \label{pp_contact}
C_{pp}^0 = \frac{\rho_{pp}^F(r_0)}{|\varphi_{pp}^0(r_0)|^2},
\ee
which  is our new relation that connects the charge distribution and the
$pp$ contact.
We recall that for calculating $\rho_{pp}^F$ we only need to know the
point-proton density $\rho_p(r)$ (or the charge distribution),
and that $\varphi_{pp}^0$ is simply calculated by
solving the two-nucleon Schroedinger equation.
The ratio of $pp$ contacts of two nuclei,
$X_1$ and $X_2$, is then given by
\be \label{pp_ratio}
\frac{C_{pp}^0(X_1)}{C_{pp}^0(X_2)}  =
 \frac{\rho_{pp}^{F,X_1}(r_0)}{\rho_{pp}^{F,X_2}(r_0)}.
\ee
where $\rho_{pp}^{F,X}$ is the uncorrelated $pp$ density
of nucleus X, with Fermi corrections. The universal two-body
functions  cancel in taking the ratio so that this contact ratio is independent of the 
model of  the nucleon-nucleon potential.

The calculations shown in Fig. \ref{40Ca} imply that this new
relation can be used to extract
the value of the $^{40}$Ca  $pp$ contact $C_{pp}^0$
using its charge distribution.
Inspecting the figure we see that the 
VMC results are well reproduced by the contact expression 
for $r\leq 1 \;\rm{fm}$ and by $\rho_{pp}^F$ for $r \geq 2\;\rm{fm}$.
In between we see a discrepancy of about 10-20\% which we attribute to the
contribution of $\ell \neq 0$ channels neglected here and to three-body correlations.
The use of the infinite nuclear-matter approximation might also have some contribution
to this difference.
We thus expect our error to be of the order of 10\%.
To get a more concrete estimate for the uncertainties in $C_{pp}^0$
we vary $r_0$ between $0.8$ fm to $1$ fm.
%The uncertainties  of the contacts are calculated by varying the value of $r_0$
%between $0.8$ fm and $1$ fm.

\begin{figure}\begin{center}
\includegraphics[width=8.6 cm]{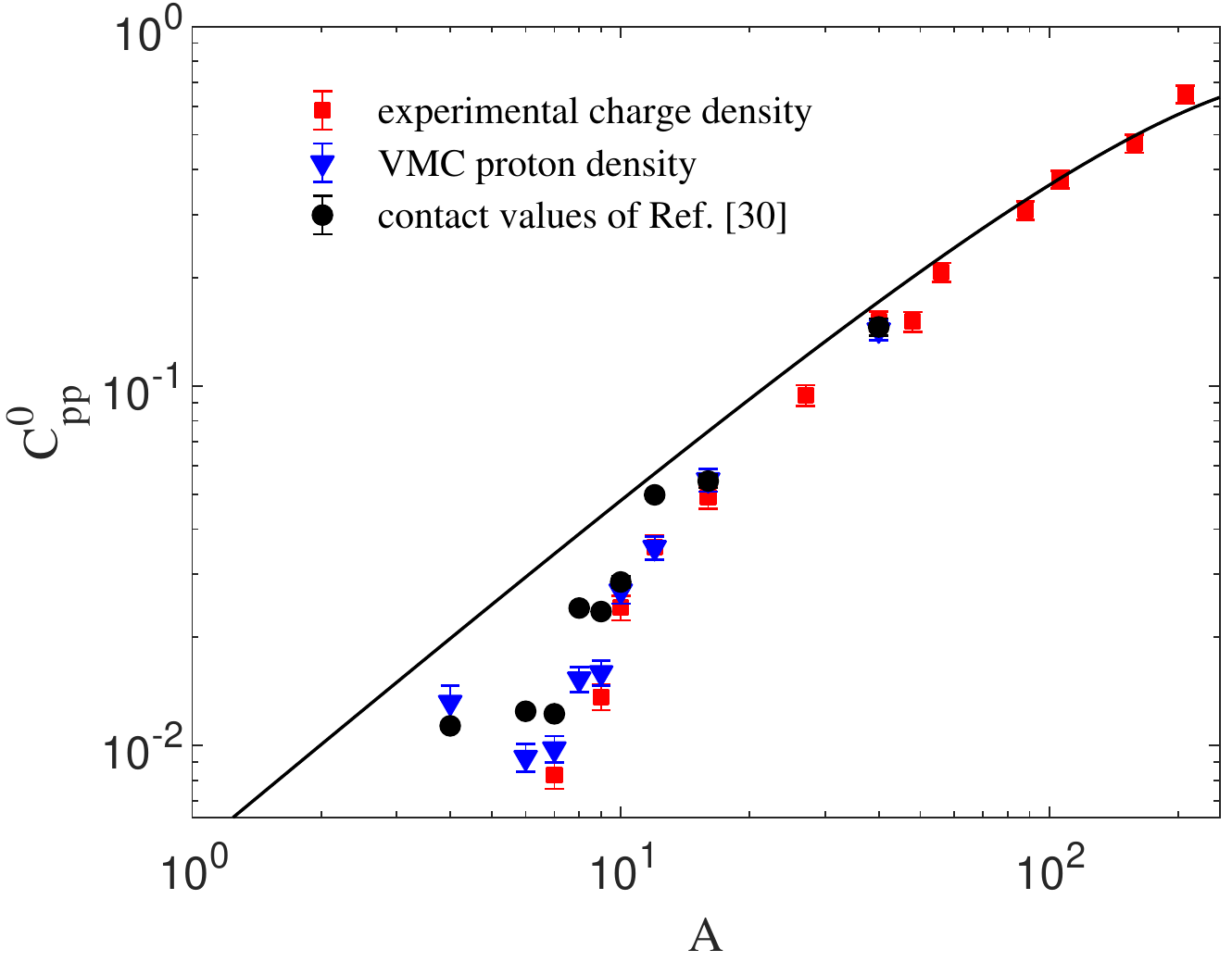}
\caption{\label{contact_values_log_A}
The $pp$ contact values as a function of A, extracted based on Eq. 
\eqref{pp_contact}, using the VMC proton density
and the experimental charge density (blue triangles and red squares, correspondingly).
Previously extracted values
of the $pp$ contacts are shown as black points
(taken from table I of Ref. \cite{WeiHen18}, without the A/2 normalization).
The black line is a fit of the form $C_{pp}^0= b Z^2/A$,
with $b=0.02$.
See the text for more details.
}
\end{center}\end{figure}

We next use Eq. \eqref{pp_contact} to extract the $pp$ contacts of different nuclei.
For nuclei up to $A=40$, we can use the VMC calculations
for the point-proton density, or the experimental data \cite{charge_denisty_parameters}.
For heavier nuclei, only experimental data is available.
Using this new method, the extracted $pp$ contacts for various nuclei,
ranging from $^4$He to $^{208}$Pb, are presented
in Fig. \ref{contact_values_log_A}, as a function of
the number of nucleons $A$, in log-log scale.
%The Fermi momentum was calculated using
%both Eqs. \eqref{kF_averaged} and \eqref{kF_central}. 
One can see that the flexibility of using either the VMC
point-proton densities or experimental charge
densities has little impact on the extracted values of the contacts. 
%The same is true for 
%the possible variation caused by using either of  two definitions of the Fermi momentum.
 Contact values of $A \leq 40$
nuclei that were previously extracted by fitting 
directly the short-distance part of the VMC two-body densities in coordinate
space \cite{WeiHen18} are also presented in the figure.
Overall good agreement is observed between
these values and the values extracted here using the charge density.
This strengthens the validity
of the relation between the nuclear contacts and the charge density.
We note that some deviations are seen for the light nuclei ($A \leq 9$).
This is expected due to the use of the nuclear matter
expression in deriving  Eq.~(\ref{large_distances_rho_pp}). Thus, 
our model for the $pp$ pair-density is best suited for application to 
medium to heavy nuclei. Notice that both $^{40}$Ca and $^{48}$Ca
are presented in the figure and have similar $pp$ contact values, given the uncertainties. 
%are presented for $Z=20$. 
%The extraction of the contact values here was done using 
%the AV18 potential for the calculation of the two-body universal functions.
The black line in the figure represents a fit of the form
$C_{pp}^0(A) = b Z^2/A$, where $b$ is a fitting parameter,
and the value of Z was estimated using the relation
\be
  Z \approx
  \frac{A}{2} \left[ 1-  \frac{a_c A^{2/3}}{4a_A}  \right]
  \approx 
  \frac{A}{2} \left[ 1-0.0075 A^{2/3} \right],
\ee
obtained from the semi-empirical mass formula
looking for the value of Z, for a given A,
that minimize the mass.
 $a_c \approx 0.711$ MeV and $a_A \approx 23.7$ MeV are the coefficients of the
Coulomb and asymmetry terms in the mass formula.
The fitted value of $b$ is 0.02.
The black line seems to describe the data well, especially
for medium size and heavy nuclei. This indicates that the $pp$ contacts, i.e.
the probability of finding a correlated $pp$ pair in the nucleus, scale like $Z^2/A$.
Similar scaling, $C_{pp}^0 \propto Z$, was postulated in \cite{WeiBazBar15a}
and can also describe these results reasonably well. A qualitative explanation of the  $Z^2/A$ behavior is that the number of pairs must be
multiplied by the probability for a proton to be at a given location, which is the inverse of the nuclear volume, i.e., the inverse of $A$ 
assuming constant density. This behavior is significantly different from
the naive combinatorial scaling of the number of $pp$ pairs.

%The black line in the figure  is a fit of the form $C_{pp}^0 = a Z$ for 
%all the values of the contacts together. The fitted value of the single free parameter is
%$a=0.0079$. This simple linear relation seems to describe the data well, especially
%for non-light nuclei, $Z \geq 6$. This indicates that the $pp$ contacts, i.e.
%the probability of finding a correlated $pp$ pair in the nucleus, scale like $Z$.

To  better understand this scaling of the $pp$ contact
values, we examine  a simple model, similar
to the Fermi-gas model, in which the nucleus is a sphere
with volume V. The proton density is just $Z/V$ inside the nucleus
radius, and vanishes outside. In this case,
for $V \rightarrow \infty$,
the integration over $R$ in $\rho_{pp}^{UC}$ just gives additional
factor of V, and we get
\begin{align} \label{rho_pp_simple_scaling}
\rho_{pp}^F(r) 
\approx 
\frac{1}{2} \frac{Z^2}{V}
 \left[1-\frac{1}{2} \left( \frac{3 j_1(k_F^p r)}
{k_F^p r} \right)^2 \right]
\end{align}
We expect this result to hold for heavy nuclei.
We can use $k_F^p = (3\pi^2 \rho_p)^{1/3}=(3\pi^2 Z/V)^{1/3}$.
%Substituting it to Eq. \eqref{pp_contact}, we obtain
%\be
%C_{pp}^0  
%= \frac{\frac{1}{2} \frac{Z^2}{V}
% \left\{1-\frac{1}{2} C^2 \left[ (3\pi^2 Z/V)^{1/3} r_0 \right] \right\}}{|\varphi_{pp}^0(r_0)|^2}
%\ee
For the volume of the nucleus we can use the approximate relation 
$V=\frac{4\pi}{3}R_0^3 A$, where $R_0 \approx 1.2$ fm.
Using these relations, it turns out that the term in the large brackets
in Eq. \eqref{rho_pp_simple_scaling} for $r=r_0$, is almost constant
for all nuclei, i.e. approximately A-independent, and equals approximately $3/5$.
Eventually, following Eq. \eqref{pp_contact}, we get
\begin{align} \label{C_pp_simple_scaling}
C_{pp}^0
\approx 
\frac{9}{40 \pi}\frac{1}{R_0^3}
\frac{1}{|\varphi_{pp}^0(r_0)|^2} \frac{Z^2}{A}
\approx 
(0.023 \pm 0.002) \frac{Z^2}{A},
\end{align}
using $|\varphi_{pp}^0(r_0)|^2 \approx 1.8311 $ fm$^{-3}$
for the AV18 potential, and $r_0 = 0.9$ fm, and estimating the error
by varying $r_0$ between $0.8$ fm to $1$ fm.
Using this simple model, we have obtained here the $Z^2/A$ scaling of the
$C_{pp}^0$ contacts, seen in Fig. \ref{contact_values_log_A}.
The numerical coefficient obtained here also agrees with
the fitted value of $b$, presented above.

As mentioned before, in principle, the point-proton density should
be used in Eq. \eqref{UC_rho_pp} and not the charge density.
A possible way for calculating the point-proton density from the experimental
charge density is described in Ref. \cite{Rocco18}. Calculating the
point density of $^{40}$Ca based on Eqs. (17) and (18) of that paper,
and assuming the neutron density is the same as the proton density,
leads to a small correction of less than 10\% in the extracted $pp$ contact of 
$^{40}$Ca. The correction for heavier nuclei is expected to be even smaller.

%Now, since $\rho_p=Z/V\propto Z/A$ is almost
%constant for non-light nuclei, we directly get
%$\rho_{pp}^F(r)  \propto Z$. Thus, using Eq. \eqref{pp_contact},
%we obtain the scaling of the $pp$ contacts observed in Fig.
%\ref{contact_values_log}
%\be
%C_{pp}^0 \propto Z.
%\ee

We have shown here that the $pp$ contacts
can be evaluated using only the charge density.
We will now present an even more surprising relation: the 
connection  between the charge density and neutron-proton SRCs.
To this end we  focus on symmetric ($N=Z$) nuclei.
For $pp$ pairs we had to consider only one channel, the spin-zero channel. 
In contrast, for $pn$ pairs we have a more complicated
situation as there are two leading SRC channels: 
the spin-zero channel
and the spin-one channel (the deuteron
channel). 
To resolve this problem we note that, due to 
isospin symmetry \cite{WeiHen18}, for symmetric nuclei
the $pp$ and $pn$ spin-zero contacts are the same. 

\begin{figure}\begin{center}
\includegraphics[width=8.6 cm]{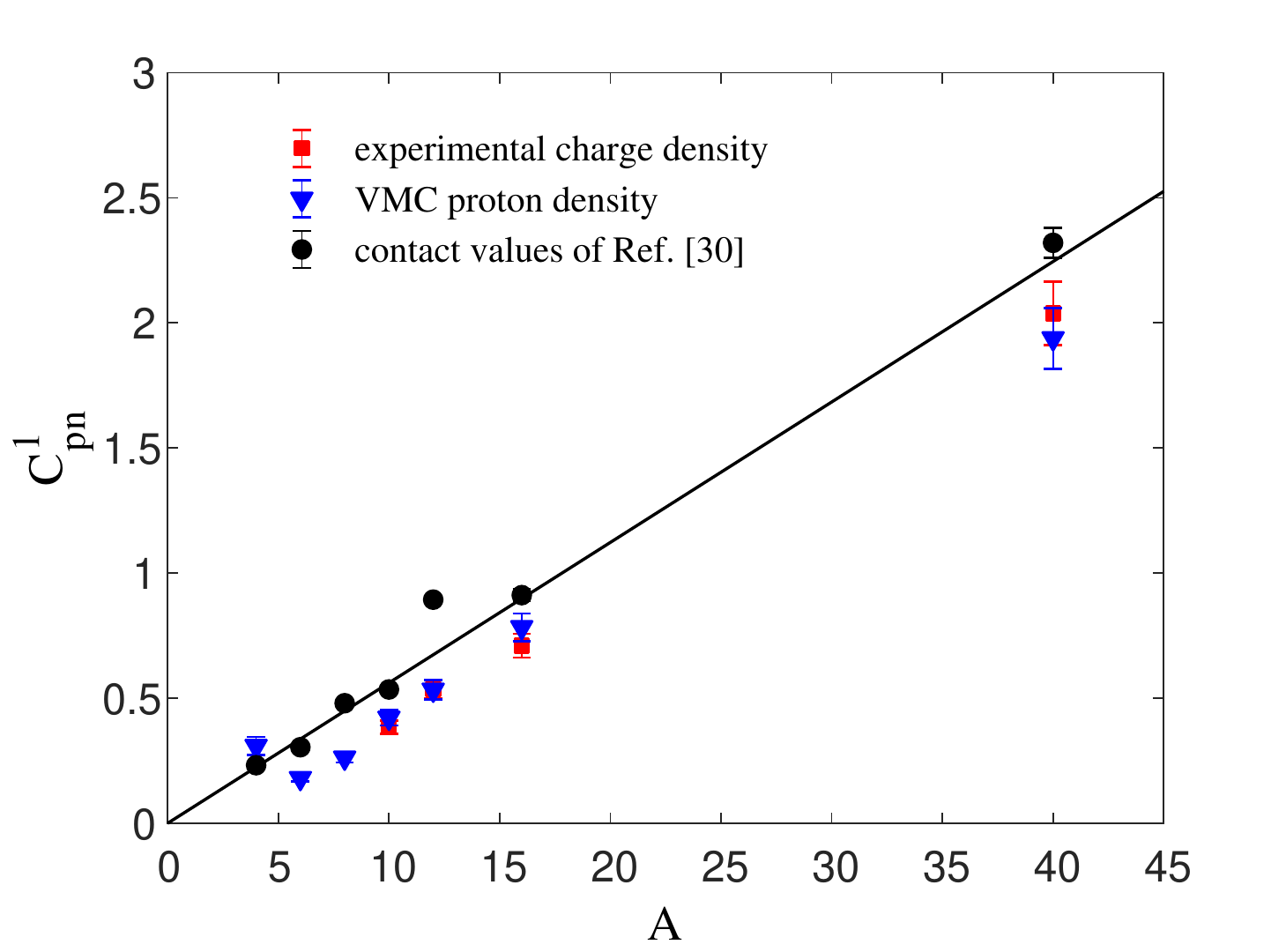}
\caption{\label{contact_values_deuteron}
The $pn$ deuteron contact values
as a function of A, for symmetric nuclei, extracted using Eq. 
\eqref{deuteron_contact_relation}, using the VMC point-proton densities
and experimental charge densities (blue triangles and red squares, correspondingly).
Previously extracted values for symmetric nuclei
are shown as black points 
(taken from table I of Ref. \cite{WeiHen18}, without the A/2 normalization).
The black line is a fit of the form $C_{pn}^1(A)=aA$,
resulting $a=0.056$.}
\end{center}\end{figure}

As before, also for $pn$ pairs we start with 
the uncorrelated two-body density given by
\be \label{UC_rho_pn}
\rho_{pn}^{UC}(\bs{r}) = \int d^3R \rho_p(\bs{R}+\bs{r}/2)
\rho_n(\bs{R}-\bs{r}/2).
\ee
The one-body neutron density $\rho_n(\rvec)$ is not as
accessible experimentally as the proton density, however
for symmetric nuclei, 
isospin symmetry implies that $\rho_n(\rvec) \approx \rho_p(\rvec)$.
%Since the one-body proton distribution is
%normalized to Z, $\int d^3r_p \rho_p(\bs{r}_p) =Z $,
%and the the neutron density to N, we get that 
%$\rho_{pn}^{UC}(\bs{r})$ is normalized to $NZ$,
%the number of pn pairs.
It follows that for symmetric nuclei 
$\rho_{pn}^{UC}(\bs{r}) \approx \rho_{pp}^{UC}(\bs{r})$.
%(including also the same normalization, $NZ=Z^2$).
Since protons and neutrons are distinguishable, there is no correction
due to the Fermi statistics here.
At small distances we use the contact relation for the $pn$ density 
\cite{WeiHen18}
\be
  \rho_{pn}(r) = C_{pn}^0 |\varphi_{pn}^0(r)|^2 + C_{pn}^1 |\varphi_{pn}^1(r)|^2.
\ee
As for the $pp$ case, we expect 
both the contact relation and the uncorrelated expression to describe
reasonably well the full $pn$ density around $r=r_0$. Thus,
by requiring only the continuity of the $pn$ density at $r_0$, we find
that
\be
   C_{pn}^0 |\varphi_{pn}^0(r_0)|^2 + C_{pn}^1 |\varphi_{pn}^1(r_0)|^2 = 
\rho_{pn}^{UC}({r_0}).
\ee
For symmetric nuclei, $C_{pn}^0 \approx C_{pp}^0$,
and also, generally, $\varphi_{pn}^0(r) \approx \varphi_{pp}^0(r)$. 
Thus, utilizing Eq. \eqref{pp_contact} we obtain
\be \label{deuteron_contact_relation}
  C_{pn}^1 = 
      \frac{\rho_{pn}^{UC}({r_0}) - \rho_{pp}^F(r_0)}{|\varphi_{pn}^1(r_0)|^2}.
\ee
This relation indicates that the ratio of two $pn$ deuteron-channel
contacts, for two symmetric nuclei, does not depend on the potential,
similar to Eq. \eqref{pp_ratio}.
Eq. \eqref{deuteron_contact_relation} can be used to extract the values of the $pn$
deuteron-channel contacts for symmetric nuclei, using only the
proton density % charge distribution
 as an input.
The results are presented in Fig. \ref{contact_values_deuteron}
for several symmetric nuclei ($A \leq 40$).
The values were extracted using both the VMC point-proton densities
and experimental charge densities.
%and both definitions of the Fermi momentum 
%(Eqs. \eqref{kF_averaged} and \eqref{kF_central}).
The values extracted using these two possibilities agree with each other for each nucleus.
The extracted values are also compared to previous values extracted by fitting to the
VMC densities directly \cite{WeiHen18}.
Fair agreement is observed. The extraction
of the $pn$ contact values using the charge density seems
to slightly underestimate the values of the contact
for $A \leq 40$.
The uncertainties of the contacts extracted here are obtained by varying $r_0$
between $0.8$ fm and $1$ fm.
The black line is a fit of the form $C_{pn}^1(A) = a A$, yielding
\be \label{pn_scaling}
  C_{pn}^1(A) =  (0.056 \pm 0.001) A.
\ee

We can analyze the $pn$ deuteron-contact scaling 
as we did for the $pp$ case. Using the same model,
we get for symmetric nuclei
$C_{pn}^1(A)  \approx (0.073 \pm 0.008) A$,
%\be
%\rho_{pn}^{UC}(r_0)-\rho_{pp}^F(r_0)  
%= \frac{NZ}{V} \left[  \frac{1}{2} + \frac{1}{4} C^2(k_F^p r_0) \right]
%= \frac{7}{10} \frac{3}{4\pi R_0^3} \frac{NZ}{A} 
%\ee
%Thus, 
%\be
%C_{pn}^1(A) 
%% \frac{21}{40\pi R_0^3} 
%%\frac{1}{|\varphi_{pn}^1(r_0)|^2}
%%\frac{NZ}{A}
% \approx (0.29 \pm 0.03) \frac{NZ}{A} \approx 
% (0.073 \pm 0.008) A,
%\ee
where we have used $|\varphi_{pn}^1(r_0)|^2 = 0.3335$ fm$^{-3}$ for $r=0.9$ fm
and $R_0 = 1.2$ fm.
This is a linear relation between $C_{pn}^1(A)$
and A, in agreement with Fig. \ref{contact_values_deuteron}.
The coefficient obtained using this simple model is larger
than the fitted value of $a$ presented above. Notice that
Fig. \ref{contact_values_deuteron} includes only
$A \leq 40$ nuclei while we expect this model to work for
heavy nuclei, as seen from the fit in Fig. \ref{contact_values_log_A}.
This can explain the difference obtained here between the two values of $a$.

The A-dependence of the deuteron-channel contact
was studied before, using a relation between the contacts
and the Levinger constant \cite{WeiBazBar15,WeiBazBar15a,WeiBazBar16}.
The Levinger constant $L$ relates the photo-absorption
cross section of a nucleus A, $\sigma_A(\omega)$,
with the same cross section for the deuteron, $\sigma_d(\omega)$,
\cite{Lev51}
\be
\sigma_A(\omega) = L \frac{NZ}{A} \sigma_d(\omega).
\ee
Here, $100$ MeV $ < \hbar \omega < 200$ MeV is the photon energy.
The idea that the photon is absorbed by
a $pn$ pair was used to show that the Levinger constant is related to
the probability to find a correlated $pn$ pair in a nucleus A
relative to that of the deuteron \cite{WeiBazBar15,WeiBazBar16}.
Using the contacts, this relation can be written as
\be
C_{pn}^1(A) = L \frac{NZ}{A} C_{pn}^1(d).
\ee
The deuteron's spin-one contact $C_{pn}^1(d)$
describes the probability to find a correlated pair
in the deuteron with momentum above $k_F$. For the AV18 potential,
and $k_F=1.3$ fm$^{-1}$, $C_{pn}^1(d)=0.0475 \pm 0.0005$.
Using the experimental estimation of $L=5.5 \pm 0.2$ \cite{WeiBazBar15}
we obtain
\be
C_{pn}^1(A) = (0.26 \pm 0.01) \frac{NZ}{A} = (0.065 \pm 0.003) A,
\ee
where the last equality holds for symmetric nuclei.
Thus, comparing to Eq. \eqref{pn_scaling},
there is agreement to within 10\% between the experimental value of the
Levinger constant and the contact values extracted
using the charge density.

In \cite{WeiHen18} the nuclear contacts were related to the
high-momentum scaling factor $a_2=(2/A)\sigma_A/\sigma_D$, 
that is extracted from inclusive 
electron scattering cross-section ratios, 
in a similar fashion to the Levinger constant.
Fomin et al. \cite{Fomin12}, evaluated $a_2$ from 
inclusive experiments carried out at 
Jefferson laboratory, and have found that 
for medium-size and heavy nuclei it is roughly a constant
$a_2 = 4.3\pm 0.3$, after correcting for center-of-mass motion.
Utilizing this value we get
$C_{pn}^1(A)=(0.085\pm 0.006)A$ for symmetric nuclei,
a value somewhat larger than the 
other extractions. For example, using the value of $C_{pp}^0$
of $^{40}$Ca, presented in Fig. \ref{contact_values_log_A}, and the
interpolated value of $a_2 \approx 4.15$ we get 
$C_{pn}^1(^{40}$Ca$) \approx 3.5$, which is more than $30\%$
larger than the value presented in Fig. \ref{contact_values_deuteron}.
The use of the value of $a_2$ without the center-of-mass
correction of Ref. \cite{Fomin12} leads to a larger discrepancy.
These discrepancies require further investigation.

To emphasize the implications of these results,
we  focus on the  example of $^{40}$Ca.
Based on Figs. \ref{contact_values_log_A} and
\ref{contact_values_deuteron}, we have $C_{pn}^1(^{40}$Ca$) \approx 2.2 $
and $C_{pp}^0(^{40}$Ca$) \approx 0.15$.
Thus, the ratio of total correlated $pn$ deuteron pairs to correlated
$pp$ pairs (with relative momentum above $k_F$) in $^{40}$Ca
is $C_{pn}^1(^{40}$Ca$)/C_{pp}^0(^{40}$Ca$) \approx 15$.
This agrees with the known dominance of correlated $pn$ pairs over
$pp$ pairs \cite{Subedi08,HenSci14}. As a result, we are led to the conclusion 
that the $pn$ dominance of SRC pairs can
be explained using only the charge distribution.
If we use the scaling of the $pp$ contacts obtained above,
and the scaling of the $pn$ contacts based on the relation
to the Levinger constant, we obtain
\be
\frac{C_{pn}^1}{C_{pp}^0} = \frac{L C_{pn}^1(d) }{b} \frac{N}{Z} 
\approx 13 \frac{N}{Z}.
\ee
This provides a prediction for the scaling of the ratio between
the amount of SRC $pn$ (deuteron) pairs and $pp$ pairs,
valid for medium-heavy nuclei, that should be checked
when sufficient experimental data will be available.
We note that the numerical factor
might be model dependent but the $N/Z$ scaling should be
model independent.

The same idea can be applied to not only the $pn$
to $pp$ ratio but also to other nuclear quantities,
such as high momentum tails and the Coulomb sum rule,
which can be described using the nuclear contacts.
On the other hand, some properties of nuclear SRCs,
such as the center-of-mass momentum distribution
of the pairs \cite{Erez18}, cannot be explored using this model.

 %=============================================================================
% Summary
%=============================================================================
%{\it Summary --}

To conclude, charge density and nuclear SRCs seem naively 
to be two unrelated aspects of nuclear systems.
Nevertheless, the use of  the generalized nuclear contact
formalism leads to the  derivation of  a direct relation between
the two. Namely, we have been able
to extract the nuclear contacts, which are proportional to the probability of finding
pairs of nucleons in a close proximity in the nucleus, using only the charge density.
The $pp$ contacts for various nuclei, and the $pn$ contacts
for symmetric nuclei, are evaluated and compared to previously
known values of the contacts, and a good agreement was observed.
Since charge densities are known for many nuclei, this provides 
a useful way for extracting SRC properties of heavy nuclei, for which
ab-initio calculations are presently almost impossible.
This new relation also shows that the ratio of $pp$ contacts, for two
nuclei,   does not depend on the choice of a  particular nucleon-nucleon
interaction. This holds also for the $pn$ contacts of symmetric nuclei.
The scaling of the $pp$ and $pn$ contacts is  also discussed
and identified, leading to a prediction regarding the
$pn$ to $pp$ ratio of SRC pairs.
The extracted values of the $pn$ contacts seem
to agree with a previous relation, connecting the Levinger
constant and the contacts, and with the known $pn$
dominance. The relation between the contacts and $a_2$
requires further investigation.

The nuclear contacts are directly related to several nuclear quantities
and reactions, such as the high-momentum tail of momentum distributions,
high momentum-transfer and energy-transfer electron-scattering experiments
sensitive to nuclear SRCs, the Coulomb sum-rule,  and the properties of nuclear matter.
%, and comparison
%with other works discussing SRCs in nuclear matter, such as Ref. \cite{nuclear_matter}.
The use of  the new relations presented in this work 
can provide predictions for such sophisticated experiments
and calculations for different nuclei using only the widely known charge distribution of each nucleus.

\begin{acknowledgments}
This work was supported by the Pazy foundation,
 by the U. S. Department of Energy Office of Science,
Office of Nuclear Physics under Award Number DE-FG02-97ER-41014,
the Batsheva de Rothschild Fellowship of  the Israel Academy of Sciences and Humanities, 
and the Shaoul Fellowship of the Sackler Institute of Tel-Aviv University.
\end{acknowledgments}

%==============================================================================
\begin{acknowledgments}
\end{acknowledgments}
%==============================================================================

%==============================================================================
\end{document}